# Notes from the Physics Teaching Lab: Diode Laser Spectroscopy at 658 nm

Kenneth G. Libbrecht[1]

Department of Physics, California Institute of Technology

**Abstract.** We describe a teaching-lab experiment that applies basic optical spectroscopy to examine the physics of semiconductor diode lasers. By using a low-power visible laser, this experiment is suitable for use in an open lab environment, where students assemble the spectrometer optics themselves as part of the project. A small holographic grating disperses light onto a camera driving a large display monitor, providing a high-resolution, real-time look at the laser's output spectrum. Observing this spectrum as a function of injection current, diode temperature, and optical feedback reveals several features of the laser's internal optical resonator and semiconductor properties. While teaching about the intrinsically quantum phenomenon of laser physics, this experiment also provides a hands-on experience that gives students a practical familiarity working with lasers, optics, and imaging technologies.

## Introduction

Creating quality physics experiments for the undergraduate teaching lab remains a never-ending challenge, in part because one would like to achieve so many independent and even conflicting goals. For example, students are often eager to race ahead to work with cutting-edge physics and technology in the lab, but teaching-lab experiments that present too much complex material too quickly can result in an unnecessarily discouraging experience. In a similar vein, high-level prefabricated experiments are good for presenting complex physics concepts, but simpler hands-on experiments are often better at teaching practical skills in electronics, optics, and other technologies that provide a foundation for working in graduate-level research labs. On the other side of the teaching-lab coin, instructors want experiments that are also modular, robust, inexpensive, and easy to maintain. Realizing all these goals simultaneously is clearly difficult, but ever-improving technology continues to present opportunities for developing better laboratory teaching tools, as we attempt to do in this paper.

Laser experiments are an excellent addition to any undergraduate teaching lab, in part because lasers are readily available room-temperature macroscopic quantum devices that are also great fun to play with. A variety of different types of lasers have long been used to demonstrate optical interferometry [1982Fen, 2013Yim, 2015Lib, 2020Gfr, 2022Alj], and tunable diode lasers are often used in atomic spectroscopy experiments that connect well to standard undergraduate quantum mechanics courses [1992Mac, 1995Lib, 1996Pre, 2006Lib, 2006Ols, 2020Bei]. In addition to teaching laser physics, it is also highly beneficial to give students some laboratory exposure to the associated hardware and optics, because lasers have become so prevalent in many areas of modern science and technology.

---

[1] klibbrecht@gmail.com

Our goal in this paper is to present a laser spectroscopy experiment that uses a low-power visible laser and is relatively easy to construct using entirely commercially available hardware, with a design that provides a student-ready hands-on laboratory experience. We will not expound on the relative merits of this experiment in comparison with other teaching-lab options, nor will we survey the field in any significant detail. Our focus here is mainly on describing this experiment in detail, providing a list of parts and approximate prices (in 2025), and expounding on the overarching principles of the underlying physics. Topics covered include general laser physics, optical resonators, grating spectroscopy, optical imaging, semiconductor physics, the transition from LED to laser behavior, and using optical feedback for laser stabilization.

## A brief laser primer

Figure 1 illustrates the basic elements of any laser system: a power source (also called the *pump*), a lasing medium, and an optical resonator. The laser medium can be thought of as a gas of simple 2-level atoms, and the main job of the power source is to put these atoms into a metastable upper excited state. As these upper-state atoms spontaneously decay, they emit photons with $h\nu = E_{2level}$, and some of this light builds up in the optical resonator, where it bounces back and forth between the two mirrors to form a standing wave. When this trapped light strikes the excited atoms, it induces *stimulated emission* that causes the atoms to emit light more readily than they would via *spontaneous emission* alone.

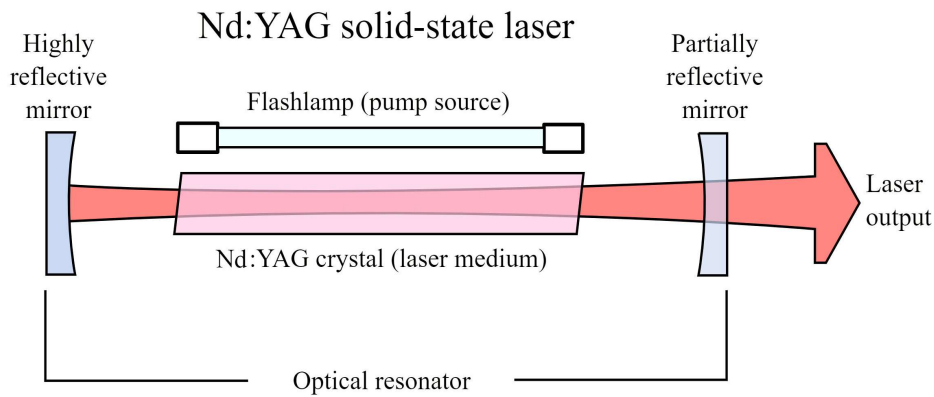

*Figure 1. This sketch illustrates the essential components of any laser system. Here a flashlamp excites neodymium ions trapped in a Nd:YAG crystal (neodymium-doped yttrium aluminum garnet), sending them into a metastable excited atomic state. Laser light builds up in the optical resonator as this light induces stimulated emission from the Nb ions at 1064 nm. Some of this light leaks out through a partially reflective mirror to provide the laser output beam. The faces of the Nd:YAG crystal are cut at Brewster's angle to reduce reflection losses within the optical resonator. Gas lasers (such as helium-neon and argon-ion lasers) have a similar construction, except the crystal is replaced by a gas-filled tube. In diode lasers, atomic energy levels are replaced by electron-hole excitations in a semiconductor crystal. (Image from https://en.wikipedia.org/wiki/Laser_construction.)*

And the photon emission process is where a key, purely quantum-mechanical effect comes into play: a stimulated-emission event sends the emitted photon into the same quantum state as the photon causing the stimulated emission. In practice, this means that the standing wave soon builds to high intensity, so nearly all the light emission is from stimulated emission, with only a small fraction being lost to spontaneous emission. When the laser reaches steady state, the output light is continuously replaced by stimulated emission in the laser medium.

The output from a laser always exhibits some kind of threshold behavior, meaning that the beam power is quite low when the pump power is low, until the output suddenly begins to increase dramatically above some threshold. The cause of this threshold behavior comes from the transition from spontaneous to stimulated emission. Looking at our toy model of a simple 2-level-atom laser in Figure 1, a low pump power means that not many atoms are driven into the exited state. These atoms decay mostly via spontaneous emission, with emission in all directions. Most of this light is lost or absorbed, so there is not much light buildup in the optical resonator. The laser output power is then just that light leaking out the partially reflecting mirror in Figure 1, which is quite low.

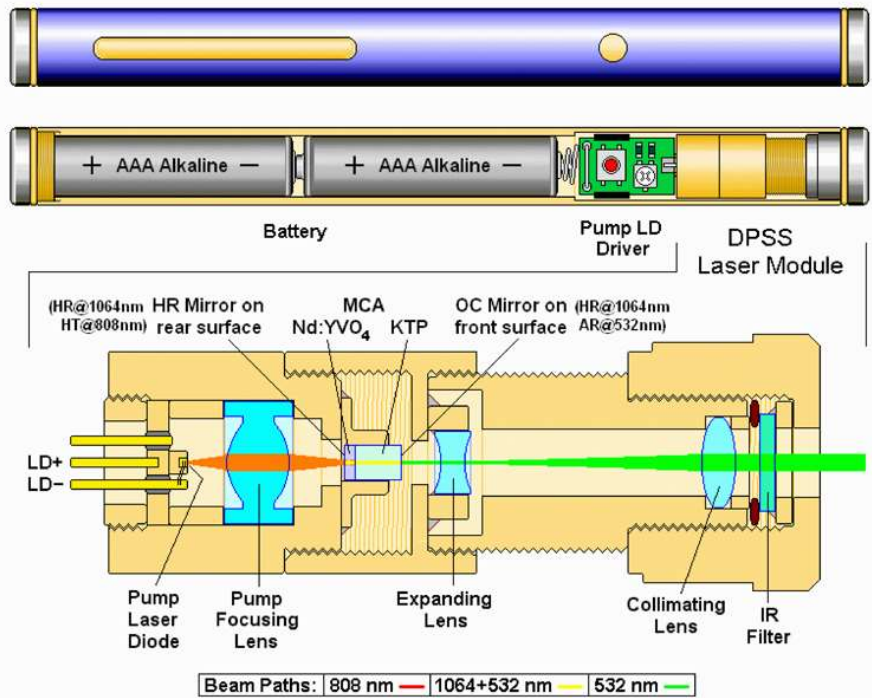

*Figure 2. A common green laser pointer illustrates the complexities that are often present in commercial laser systems (DPSS = diode-pumped solid-state; MCA = multiple-crystal assembly). Here a near-IR semiconductor diode laser serves as the power source, optically "pumping" the ions in a neodymium-doped yttrium orthovanadate (Nd:YVO4) crystal. Note that the input mirror (a multilayer dielectric coating) on the crystal allows 808-nm pump light to enter the crystal but reflects the 1064-nm laser light inside the crystal. The output beam then passes through a KTP (potassium titanyl phosphate) nonlinear crystal that converts pairs of 1064-nm photons into 532-nm photons, thus "frequency doubling" the infrared light into green light. Finally, the beam exits through an IR filter that removes any residual (invisible) 1064-nm light. Someday this complex assembly will likely be replaced by a single green diode laser, but such a simple solution does not yet exist. (Image from [2011Gal].)*

As the pump power is increased, more atoms are driven into the excited state, and the light buildup in the resonator begins to become substantial. At some point, there is a rather abrupt transition (the laser threshold) when the atomic decay becomes dominated by stimulated emission. This stimulated emission is driven by the standing wave in the cavity, so those photons are not lost like those from spontaneous emission. Above threshold, essentially all the atomic emission is into the cavity mode, where it contributes directly to the laser output power. Below threshold, the production of light is quite inefficient: a lot of input power is needed to produce little output power in photons. Above threshold, the conversion of electrical power to optical photons is much higher. The overall conversion efficiency of electrical power to laser photons varies greatly with laser type, from perhaps 0.1% for gas lasers like the popular Helium-Neon laser to above 50% for some semiconductor diode lasers.

Laser technology is a fascinating mix of quantum optics, atomic physics, semiconductor physics, and materials science, giving laser experiments a lot of pedagogical value in undergraduate teaching labs. Popular laser types include gas lasers, solid-state crystal lasers, and semiconductor lasers. Even something as commonplace as a green laser pointer involves a remarkable mix of different technologies, as shown in Figure 2. Laser physics continues to be a very active area of current research, driven by countless applications in telecommunications, display technology, spectroscopy, LIDAR, and many other areas.

## Semiconductor diode lasers

Semiconductor diode lasers rely on stimulated emission like all lasers, but the gain medium involves electron-hole pairs rather than atomic transitions. In a nutshell, sending a current through a semiconductor with p-type and n-type doping (see Figure 3) causes a recombination of electrons and holes that emits light, making a light-emitting diode (LED). The vast subject of semiconductor physics is certainly beyond the scope of this small paper but suffice it to say that the wavelength of the emitted light is largely determined by the bandgap within the semiconductor.

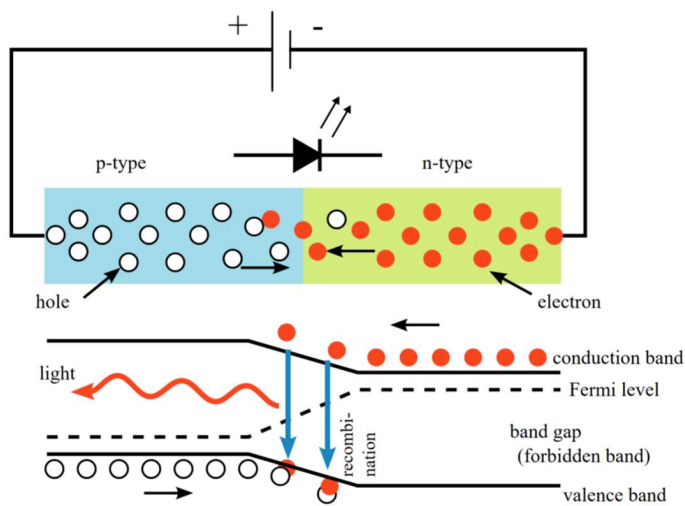

*Figure 3. An LED consists of a p-type semiconductor (with an electron depletion, forming "holes" in the valence band) bonded to an n-type semiconductor (with an electron excess populating the conduction band). When a current is driven through the junction, the electrons and holes recombine, emitting light. The photon energy is essentially equal to the energy band gap in the semiconductor. (Image from https://en.wikipedia.org/wiki/Light-emitting_diode_physics.)*

Creating a diode laser involves encasing the semiconductor gain medium within a microfabricated optical waveguide/resonator made using transparent semiconductor materials with different refractive indices. Everything is fabricated on a chip and driven by sending an "injection current" through the diode. Once light leaves the tiny laser cavity, an external lens is needed to collimate the strongly diverging laser light into a narrow laser beam. (Alternatively, the semiconductor cavity output could be immediately coupled into an optical fiber.)

Our primary focus in this paper will be on using diode-laser technology to make tunable, single-frequency laser beams that we can eventually be used for atomic spectroscopy and other applications. This focus on spectroscopy is interesting in its own right, but it also provides some useful pedagogy, as it demonstrates many aspects of laser physics with a relatively simple laboratory setup.

Our story begins with a diode laser in a can, like the example shown in Figure 4. The chip itself is roughly 1mm in length, and of course this diagram glosses over the semiconductor physics, the optical layout, the electrical contacts, and all the vital details that make this device actually work. Unfortunately, bare diode lasers like this have a host of undesirable properties:

- They have relatively low output power.
- The output wavelength is often ill-determined.
- The wavelength is difficult to tune.
- The laser beam quality is relatively poor.

On the other hand, diode lasers are quite inexpensive, and the output power can easily be modulated at GHz frequencies. The low price is an especially attractive feature, of course, so people have found clever ways to make diode lasers work in an ever-increasing range of technological applications ... including those in many areas of modern experimental physics.

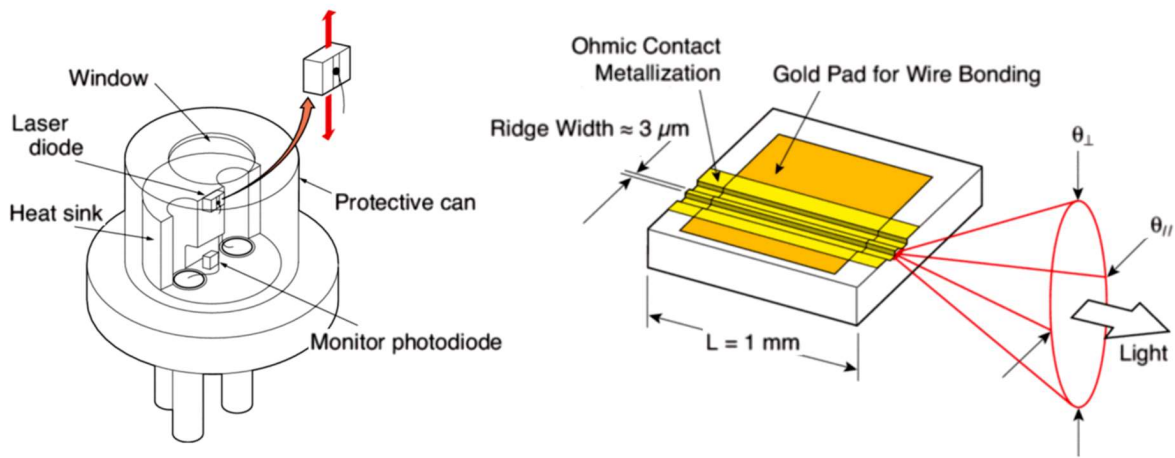

*Figure 4. (Left) Individual diode lasers are often mounted in several-mm-diameter "cans" that include the laser itself on a heat sink, a protective window for the output beam, and perhaps a photodiode to monitor the optical power. (Right) The laser optical cavity is etched into a block of semiconductor material, using different materials to confine the cavity modes. The micron-scale geometry of the semiconductor cavity results in a strongly diverging elliptical output laser beam. (Images from https://www.newport.com/t/laser-diode-technology.)*

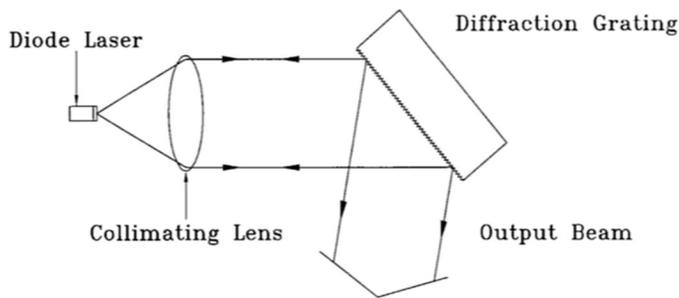

*Figure 5. This illustration shows the basic parts of the grating-stabilized diode laser (figure from [1995Lib]). Most of the laser light exiting the semiconductor is collimated and then undergoes a simple reflection off the Diffraction Grating (the m=0 Output Beam). But some light is diffracted from the grating (the m=1 beam) and goes back into the semiconductor. The grating surface and the two internal cavity mirrors form a compound system including an "internal" cavity (inside the semiconductor) and an "external" cavity (including the grating).*

In this teaching-lab experiment, part of our focus is on a device called an External-Cavity Diode Laser (ECDL) [1991Wie, 1992Mac, 1995Lib, 2000Con, 2007Car, 2015Bre], which involves using a diffraction grating to send a small amount of controlled optical feedback into a bare diode laser, as illustrated in Figure 5. The backward reflection from this grating produces a second, frequency-selective optical resonator, and this optical feedback is a popular method for making diode lasers more suitable for a variety of spectroscopic applications. As we will see below, the external grating stabilizes the laser output into a single longitudinal mode of the resonator, allows better frequency tunability, and even reduces the laser linewidth. The ECDL is widely used in experimental physics research labs, plus it provides a fascinating bit of laser technology that can be used in the teaching lab to introduce students to some basic concepts in quantum optics.

To begin to understand what is happening with this ECDL setup, we define a fictitious "generalized gain" (following [1995Lib]) that has the components shown in Figure 6. The curves in this figure provide a greatly simplified picture of a very complex phenomenon, but they can be thought of as the probability that lasing occurs at a particular frequency. The individual curves describe different physical effects, and collectively they contribute to the laser's operation. Semiconductor laser technology is an enormous field, of course, so our modest goal here is to examine just this basic, qualitative toy model of the underlying physics.

The *Semiconductor-medium* curve in Figure 6 illustrates the semiconductor material itself, and the broad peak in this function occurs at the electron-hole recombination energy. This curve is essentially the spectrum of a bare LED, which peaks at some wavelength and has spectral width of roughly 1-2 nanometers – much broader than the other curves in Figure 6. The only way to change this curve appreciably is by changing the semiconductors materials being used, although the peak wavelength will shift with temperature. There is a whole industry devoted to the *Semiconductor-medium* curve, of course, and one can purchase diode lasers with a large (but finite) selection of different wavelengths.

The *Internal cavity* curve comes from the optical resonator (a.k.a. optical cavity) inside the semiconductor material, and here the peaks correspond to standing waves (a.k.a. resonant modes) inside the resonator. As mentioned above, all lasers involve optical resonators, and a diode laser is far more likely to operate when the wavelength equals one of the resonant wavelengths of the internal cavity.

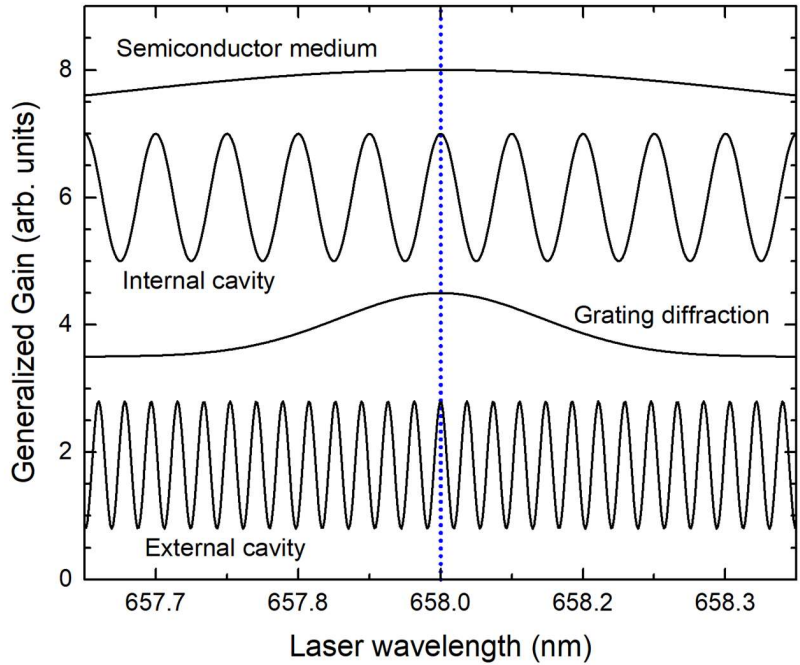

*Figure 6. Tuning an External-Cavity Diode Laser (ECDL) to a specific wavelength $\lambda_o$ (blue dotted line) requires a careful alignment of its various components. First (top curve), the semiconductor temperature is adjusted so lasing near $\lambda_o$ is preferred. Second (2nd curve), the semiconductor temperature and current are further adjusted so the internal cavity has a resonance at $\lambda_o$. Third (next curve), the grating angle is positioned so light at $\lambda_o$ is retroreflected back into the semiconductor laser. And finally (bottom curve), the grating position is adjusted so the external cavity also has a resonance at $\lambda_o$. Achieving stable laser operation at $\lambda_o$ thus requires the simultaneous setting of the laser temperature, laser injection current, grating angle, and grating position. Note that the vertical axis is not to scale in this diagram, and the curves are only meant to illustrate the underlying physical principles.*

With a bare diode laser like that in Figure 4 (with no external grating) the *Semiconductor-medium* and *Internal cavity* curves in Figure 6 are all that matter, so we begin our discussion there. Moreover, the physics is generally easier to understand when we can examine a specific test case in detail, so let us now go into the lab and set up our laser experiment. This is, after all, one of our principal pedagogical goals in a physics teaching lab – to connect abstract theoretical concepts to real-world laboratory measurements.

## Diode laser spectroscopy

Figure 8 shows the first-stage optical layout for this experiment, which provides a good introduction to optical spectroscopy using a simple diffraction grating. We have found that the overall student experience goes more smoothly if we pre-align the optical elements on the left side of this layout, leaving the grating and lens alignments to the students. To this end, we assemble the laser, beamsplitter, slit, prism, and camera on a small breadboard as shown in Figure 7, bolting these items in place with Torx-head bolts that students cannot adjust. The lens and grating are left to the students to align, attaching these elements to the table using standard hex-head bolts.

Because many of our students enter the lab with no previous optics experience, the handout walks them through the alignment procedure in detail. One begins by removing the lens and placing just the grating, aligning it to produce a streak of laser light on the camera, as illustrated in the top image in Figure 9a. Once this image has been obtained, the lens is then placed and the position of the lens (perpendicular to the optical axis) is adjusted so the light streak appears at about the same position on the camera as it was without the lens. This procedure ensures that the beam passes through the center of the lens without being deflected, as it should.

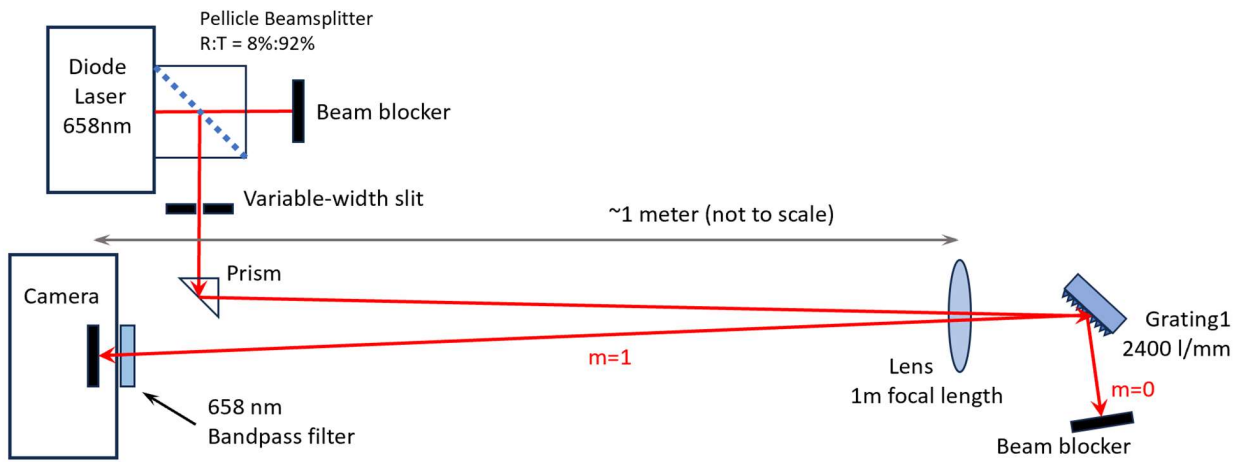

*Figure 8. This sketch shows the beginning optical layout for the laser-spectroscopy experiment. Laser light passes through a variable-width slit and is dispersed by a holographic grating, and the lens focuses the slit onto the camera sensor. In our realization of this setup, the optical elements on the left side of the diagram are all pre-aligned, while students place the lens and grating to produce the desired spectral image. Figure 7 shows some setup in the lab, and some hardware details are provided in Appendix 1.*

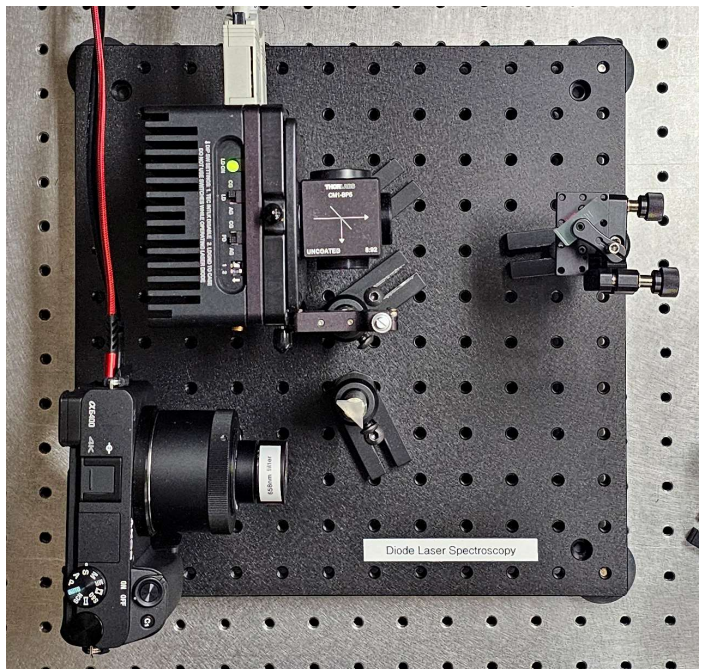

*Figure 7. This photo shows several of the components in Figure 8 aligned on a 12-inch breadboard with holes on a 1-inch pattern. Appendix 1 lists the different components with 2025 prices.*

Next the lens axial position and the slit width can be adjusted to yield the spectrum shown in Figure 9b, which appears when the laser current is below threshold. The sharpest spectral image occurs when diffraction from the slit expands the transmitted beam so it just fills the grating surface. This gives the best wavelength resolution for this grating spectrometer, given by

$$\delta\lambda_{res} \approx \frac{m\lambda}{N} \approx 0.02 \text{ nm} \tag{1}$$

where $m = 1$ is the diffraction order, $\lambda = 658$ nm is the laser wavelength, and $N \approx 36000$ is the number of illuminated grooves on the grating (2400 line/mm times ~15mm). The precise lowest value of $\delta\lambda_{res}$ is not easily calculated, given that the intensity of the laser beam varies with position on the grating, but $\delta\lambda_{res}$ is readily optimized by tweaking the slit width and the lens axial position while observing the spectrum on the camera monitor. A laser spectrum like Figure 9b can be obtained with a modest amount of alignment effort.

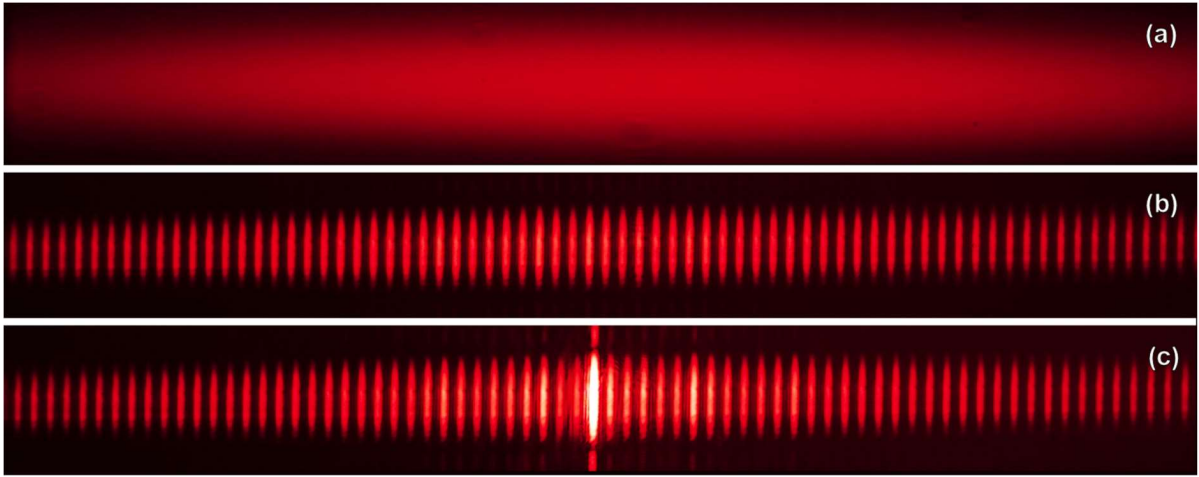

*Figure 9. This composite image shows several laser frequency spectra obtained using the setup in Figure 8. The top spectrum (a) was taken without the lens in place, so the slit is not focused on the camera. Spectrum (b) includes the lens, showing the cavity mode structure below the laser threshold. And spectrum (c) was taken above threshold, when most of the laser power was concentrated into a single dominant cavity mode.*

Looking at the physics underlying these spectra, the large-scale width of this red streak of light in Figure 9 corresponds to the *Semiconductor-medium* in Figure 6, and this curve is roughly the same as the spectral emission of an LED made from the same semiconductor material. From the spectrum in Figure 9a we measure a width of about $\delta\lambda_{medium} \approx 2$ nm for the *Semiconductor-medium* curve, or equivalently $\delta\nu_{medium} \approx 1.4$ THz.

If we assume that a thermally broadened density of states around the band gap is responsible for the *Semiconductor-medium* curve, then we might expect $\delta(h\nu_{laser}) \approx kT$, which predicts a linewidth of roughly $\delta\lambda_{medium} \approx (kT/h\nu_{laser})\lambda_{laser} \approx 9$ nm. Although this is larger than $\delta\lambda_{medium} \approx 2$ nm observed Figure 9a, this simple calculation does suggest that thermal broadening is a significant player in determining $\delta\lambda_{medium}$. Indeed, experiments show that cooling LEDs to cryogenic temperatures can greatly reduce the spectral emission of the emitted light. The full semiconductor story is quite complicated, however, and there is a large literature investigating the physics of light-emitting diodes and semiconductor lasers. For the purposes of this short experiment, we simply point out that the spectral width quantified by $\delta\lambda_{medium}$ leads one toward a large and important area of applied physics research.

## Internal cavity modes

Going beyond the *Semiconductor-medium* curve, the regimented stripes seen in Figure 9 show the resonant modes of the laser's internal cavity, illustrated by the *Internal cavity* curve in Figure 6. The mode spacing can be calculated by realizing that each cavity mode is a standing wave with an integer number of half-wavelengths filling the diode's internal resonator. We write

$$j\frac{\lambda_j}{2} = nL_{cavity} \tag{2}$$

where $n$ is the index of refraction inside the semiconductor material, $L_{cavity}$ is the length of the resonator cavity, $\lambda_j$ is the optical wavelength of the $j^{th}$ cavity mode, and $j$ is an integer index. It follows that the spacing between cavity modes is

$$\delta\lambda_{cavity} \approx \frac{\lambda^2}{2nL_{cavity}} \tag{3}$$

or equivalently

$$\delta\nu_{cavity} \approx \frac{c}{2nL_{cavity}} \tag{4}$$

when $\lambda_{laser}/L_{cavity} \ll 1$.

Students can measure $\delta\lambda_{cavity}$ directly by combining the spectral information in Figure 9b with the grating equation and the overall geometry of the optical layout. Unfortunately, we find that many of our students have had little to no experience handling imaging data in a scientific context, mainly because we have not taught them this particular skill in any of their other coursework. Moreover, software tools have become quite expensive while many university site licenses have diminished in scope, further exacerbating this pedagogical challenge. One solution is to provide a computer in the teaching lab that has suitable image-processing software already installed, but this approach does not promote student independence by giving them a tool they can apply outside the lab environment.

Another approach is to take advantage of the recent emergence of powerful new AI tools (as of 2025), and Appendix 2 shows an example of how ChatGPT-5 can be used to analyze these spectral data. The analysis is mostly straightforward once the cavity-mode spacing $\delta x_{mode}$ on the camera sensor has been determined, but students often miss that the angular separation of the cavity modes (as seen by the grating) is given by $\delta\theta = \delta x_{mode}/f_{lens}$, where $f_{lens} = 1$ meter is the focal length of the lens in Figure 8. Somewhat counterintuitively, this value of $\delta\theta$ is independent of the distance between the lens and the grating.

Putting in the numbers as shown in Appendix 2, we obtain

$$\delta\lambda_{cavity} \approx 0.0770 \pm 0.0002 \text{ nm} \tag{5}$$

or equivalently

$$\delta\nu_{cavity} \approx 53.37 \pm 0.15 \text{ GHz} \qquad (6)$$

Comparing with Equation (1), we see that $\delta\lambda_{cavity} \approx 4\delta\lambda_{res}$, so it makes sense that the individual cavity modes are nicely resolved in Figure 9b. We note that the measurement in Equations (5) and (6) do not include the slight change in these numbers with mode frequency. And ChatGPT, like most automated fitting routines, produces somewhat optimistic measurement uncertainties.

We sacrificed one of our HL6501MG diodes to measure the length $L_{cavity} = 0.787 \pm 0.026$ mm of the internal cavity (see Appendix 1), and Equation (4) then gives the cavity index of refraction $n \approx 3.6$. This is well inside the range of AlGaInP diode lasers, although these are complex heterostructures made from different materials, so each diode laser part number will have its own effective internal-cavity index of refraction.

Finally, we note that a single cavity mode dominates the spectrum in Figure 9c, which happens when stimulated emission becomes more important than spontaneous emission during electron-hole recombination. In an ideal laser scenario, the chosen cavity mode would be at the peak of the *Semiconductor-medium* curve in Figure 6, where the overall generalized gain in this toy model is the highest. Once this mode begins to dominate inside the cavity, stimulated emission "steals" available excited electron-hole pairs from the other cavity modes, and soon most of the light inside the cavity is in a single cavity mode. Thus, stimulated emission produces a winner-take-all *mode-selection instability* in the laser operation: once one cavity mode grows in amplitude, it gobbles up nearly all the available optical energy in the system, depriving the other available modes.

Note that the spectrum in Figure 9c is a bit misleading because the brightest cavity mode strongly saturates the imaging sensor. Once the bright pixels reach the maximum digital level of 255 in the 8-bit camera system, they cannot register brighter values. We can see this by comparing images taken using different camera shutter speeds, and Figure 10 shows an example of this. With a long exposure (1/40th second in this demonstration), the dominant mode is strongly saturated while the non-lasing modes are easily visible in the spectral image and in a plot of the pixel brightness values. Producing a spectral image with no saturated R pixel values required a substantial 100x reduction in exposure time (to 1/4000th second, the camera limit), and our analysis of this image revealed that >98 percent of the optical power was in the dominant cavity mode.

For the case of a real (non-ideal) diode laser, it is often the case that the dominant laser emission is not confined to a single cavity mode. For a typical diode laser like the HL6501MG, single-mode operation can be observed at some values of temperature and injection current, but multi-mode operation is commonly observed at other settings. Some of this arises from imperfections in the semiconductor devise fabrication, so each individual diode will behave slightly differently from all others. Single-mode operation is obviously desirable if one wants a frequency-tunable light source for laser spectroscopy applications, but multi-mode operation may be fine if all you need is a bright light source. Here again, we will not be focusing much of our discussion on the all the physical origins of these different diode-laser behaviors, or how they might be managed using a variety of fabrication tricks.

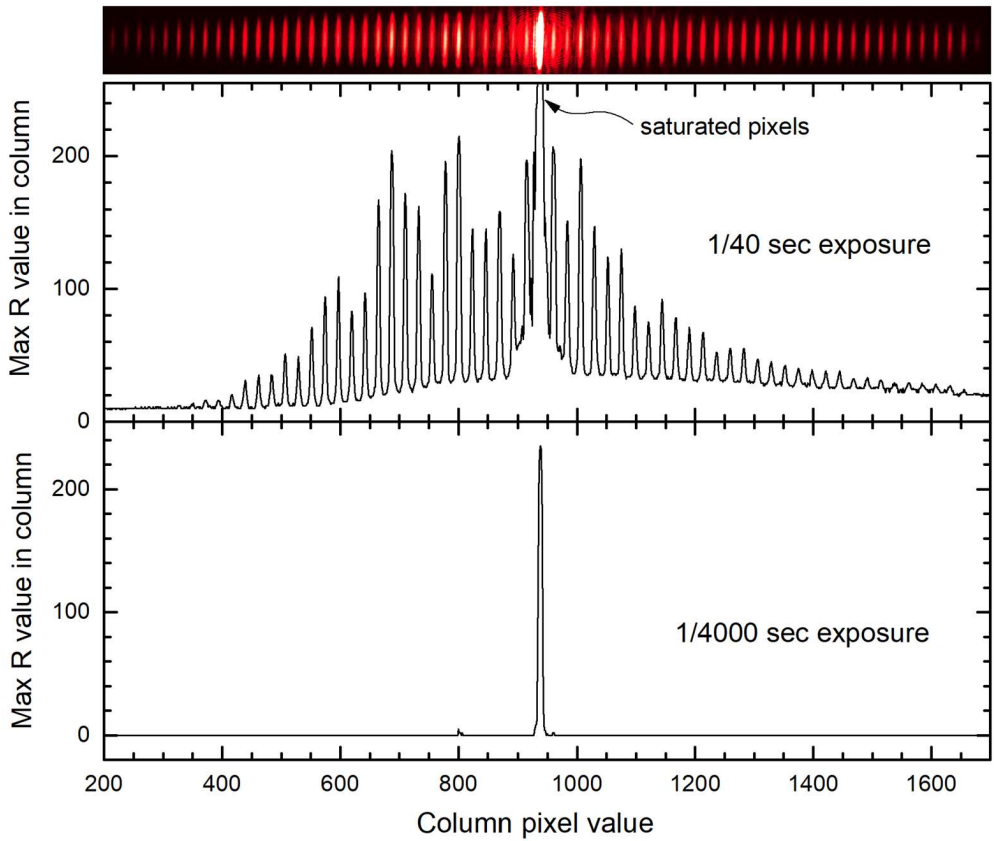

*Figure 10. The image slice at the top of this figure shows a laser spectrum with an injection current far above threshold (65 mA at 21.5 C in this case), like that seen in Figure 9c. Most of the laser's optical power comes from a single cavity mode, which is the brightest feature in the spectrum. The graph below shows the maximum R pixel value (in the RGB image) in each column of the image, revealing that the brightest pixels are saturated at a digital brightness level of 255. Reducing the camera exposure time by 100x gives the unsaturated spectrum in the lower panel, from which we find that >98 percent of the laser power is in the dominant cavity mode. This exercise illustrates that an image with saturated pixels can be quite visually misleading.*

# Temperature dependence

Observing the diode laser spectrum as a function of temperature provides additional insights into the relevant semiconductor physics of these devices. Figure 11 shows a series of spectra taken at a constant injection current of 65 mA, which is well above the laser threshold. A close look at these data reveals that the cavity modes all shift down smoothly in frequency (up in wavelength) as the temperature increases, while the dominant lasing mode shifts erratically to longer wavelengths. Acquiring these data is straightforward, making it a good student exercise.

To explain the physics behind these behaviors, consider the model in Figure 6 with the additional assumption that both the *Semiconductor medium* curve and the *Cavity mode* curve shift with temperature. The peak in the *Semiconductor medium* curve reflects the semiconductor bandgap, and it should not be surprising that this material property changes slowly with temperature. Likewise, Equation (2) indicates that the wavelength $\lambda_j$ of a specific cavity mode depends on the length and index of refraction of the semiconductor cavity, so again we would expect the *Cavity* *mode* curve to shift slowly with temperature. Moreover, there is no reason to expect that these two curves would shift at exactly the same rate with temperature.

A careful look at Figure 11 confirms that the below-threshold cavity modes shift slowly and uniformly with temperature, reflecting the slow changes in the laser cavity length and index of refraction with temperature. Figure 12 shows the same data as in Figure 11 but includes a fit line that gives a cavity-mode temperature coefficients of $d\lambda_{mode}/dT \approx 0.038$ nm/C.

The peak of the *Semiconductor medium* curve also shifts uniformly with temperature, and Figure *12* gives a temperature coefficient of $d\lambda_{laser}/dT \approx 0.17$ nm/C. At any given temperature, the dominant laser mode (if there is one) will tend to occur near the peak of the *Semiconductor medium* curve, as we discussed above. Because the *Semiconductor medium* curve shifts faster than the *Cavity mode* curve, the dominant laser frequency must occasionally "hop" between different cavity modes as the temperature changes.

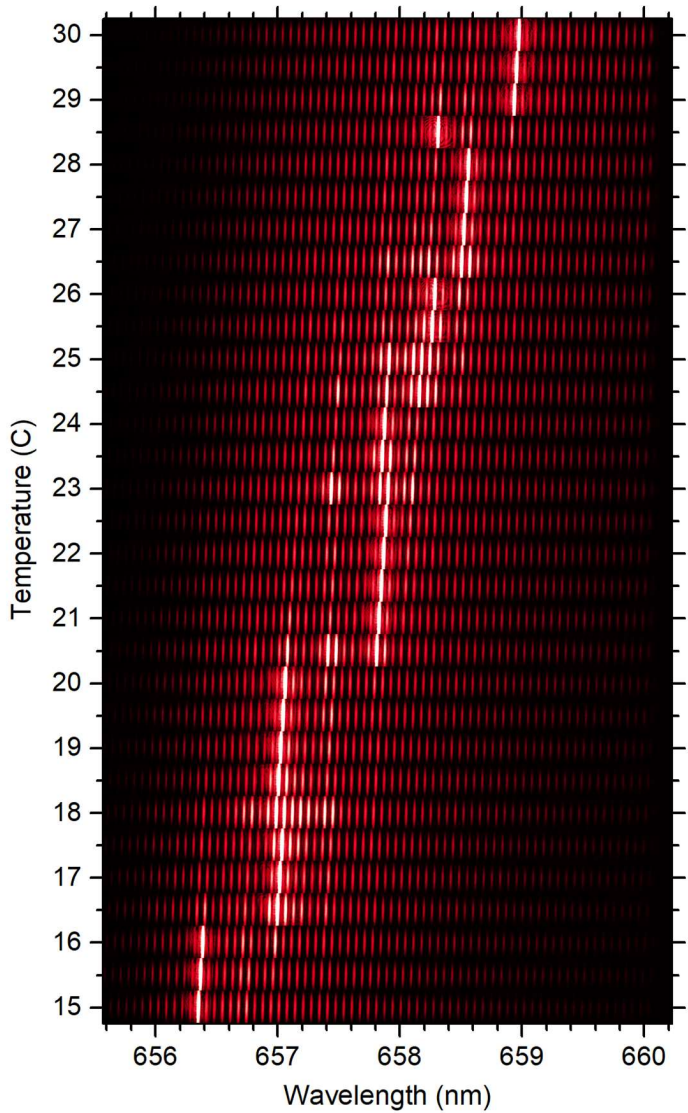

*Figure 11. This composite image shows a series of laser spectra taken at different temperatures, at a fixed injection current of 65 mA. The wavelength generally shifts upwards as the temperature increases, although not following a simple or even monotonic function. Note also that single-mode behavior dominates at some temperatures (for this particular value of the injection current), while multi-mode behavior is prevalent elsewhere.*

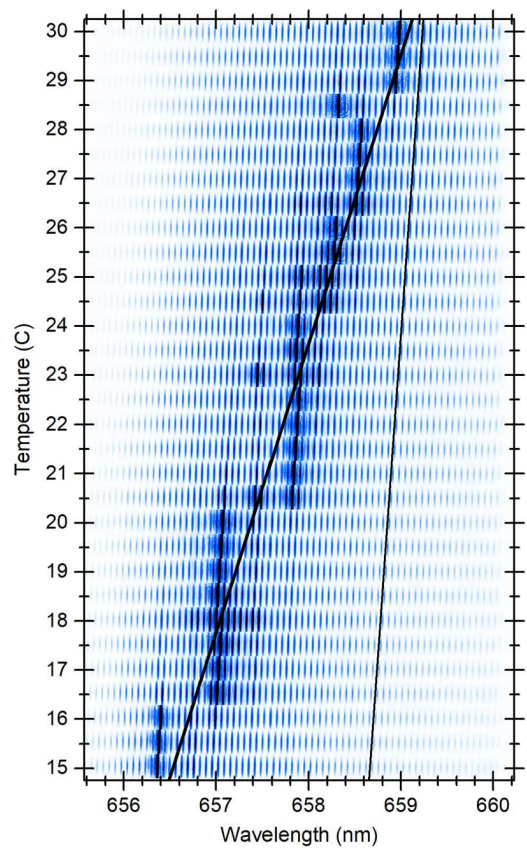

Figure 12. This inverted version of Figure 11 includes two fit lines showing slopes of $d\lambda_{mode}/dT \approx 0.038$ nm/C for a single cavity mode and $d\lambda_{laser}/dT \approx 0.17$ nm/C for the dominant lasing mode. The first slope corresponds to one of the Internal cavity peaks in Figure 6, while the second slope corresponds the peak of the Semiconductor medium curve. The fact that these two slopes are not equal makes mode hopping as a function of temperature inevitable in this diode laser.

In an ideal diode laser, these mode hops might occur at regular intervals as the two curves both move uniformly with temperature at different rates. Real devices have imperfections, however, so the mode hopping seen in Figure 11 is somewhat irregular and unpredictable. Some cavity modes exhibit stable laser operation over temperature intervals of a degree or two, while some cavity modes will not be dominant at any temperature. Also, clean single-mode behavior is seen at some temperatures (at this particular injection current), while multi-mode behavior dominates at other temperatures. The exact characteristics of these spectra will change with injection current and will also be different for each specific device.

In contrast to our spectral data, Figure 13 was taken from the HL6501MG manufacture's specifications, and this plot is somewhat misleading in that it suggests that the laser frequency is a generally well-defined function of temperature. While this might be true in some average sense, the full spectral information in Figure 11 paints a somewhat different story, revealing that multi-mode laser operation is quite common.

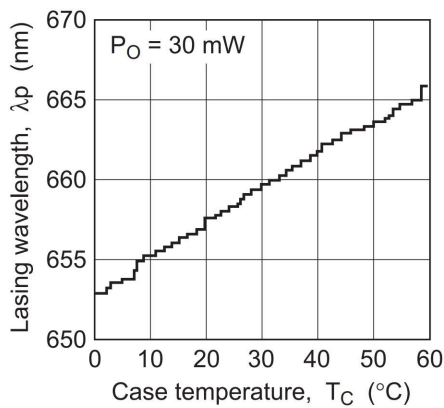

Figure 13. This plot from the HL6501MG data sheet shows ostensibly the same information as that in Figure 11, indicating an overall nearly constant dλ/dT punctuated by occasional mode hops. But the full spectral information better illustrates the occurrence of single-mode and multi-mode behaviors.

## Threshold behavior

Figure 14 shows measurements of the laser output power as a function of injection current, illustrating an extrapolated lasing threshold at 43.8 mA in this example. Again, acquiring these data is a straightforward student exercise. The linear behavior above threshold is typical for laser operation, illustrating a high efficiency turning electrical power into laser photons in this regime. The high efficiency results in part because stimulated emission from electron-hole recombination yields photons in the main laser mode, and the light escaping from the front of the cavity is all focused onto the photodetector. Below threshold, the efficiency is much lower because most of the light produced by spontaneous emission is not confined to the cavity modes and does not reach the photodiode. One highly beneficial property of lasers is that the cavity modes yield a narrow collimated beam, so the laser light can be guided to where you want it.

We used a commercial laser power meter to produce the data in Figure 14, as this specialized instrument is calibrated to measure optical power at a specified wavelength. A power meter is overkill for this teaching-lab experiment, however, and an uncalibrated photodetector (such as the Thorlabs PDA100A2) is more than adequate to produce similar data. One could also use Figure 14 to calibrate whatever photodiode you are using, as the slope above threshold will be nearly the same for all HL6501MG diode lasers. The threshold current can vary substantially with temperature and from device to device, however, so Figure 14 cannot be used to accurately predict the optical power as a function of injection current for all devices.

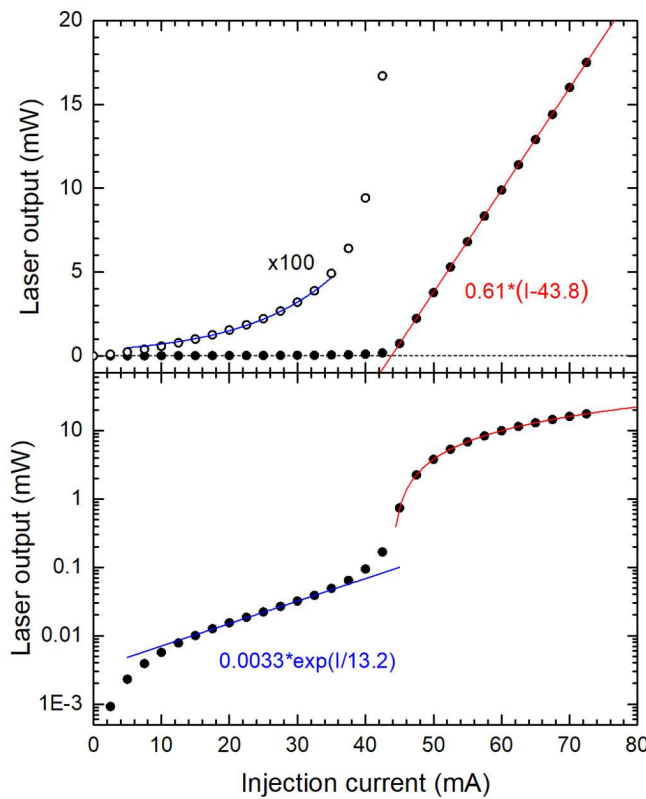

*Figure 14. The solid points show the measured laser power as a function of the injection current, illustrating an extrapolated laser threshold of 44.8 mA. Above this threshold, the optical power increases linearly with current. The open points show the same data multiplied by 100, illustrating an exponential rise in optical power as a function of injection current below threshold. The two panels show the same data and curves in different plot formats.*

## Current dependence

Figure 15 shows several laser spectra at fixed temperature for different injection currents. The behavior of the lasing modes is somewhat erratic above threshold, much like we saw in the temperature-dependent spectra above. However, the nearby non-lasing LED-like modes are quite well behaved, yielding the wavelength measurements also shown in Figure 15. A linear trend appears above threshold, and a fit to the data gives $d\lambda_{mode}/dI \approx 5.3$ nm/Amp.

This current dependence can be mostly explained by the cavity temperature increasing with increasing injection current, even when the temperature of the heatsink is fixed. In this picture, the value of $d\lambda_{mode}/dI$ increases above threshold because there is a much higher concentration of optical power (and thus heating) within the cavity itself. Comparing $d\lambda_{mode}/dI \approx 5.3$ nm/Amp with the above measurement of $d\lambda_{mode}/dT \approx 0.038$ nm/C suggests that the cavity heating above threshold can be described by $dT/dI \approx 140$ C/Amp. At the highest current we typically operate, $dI$ is about 30 mA above threshold, giving a temperature rise of about 4 C.

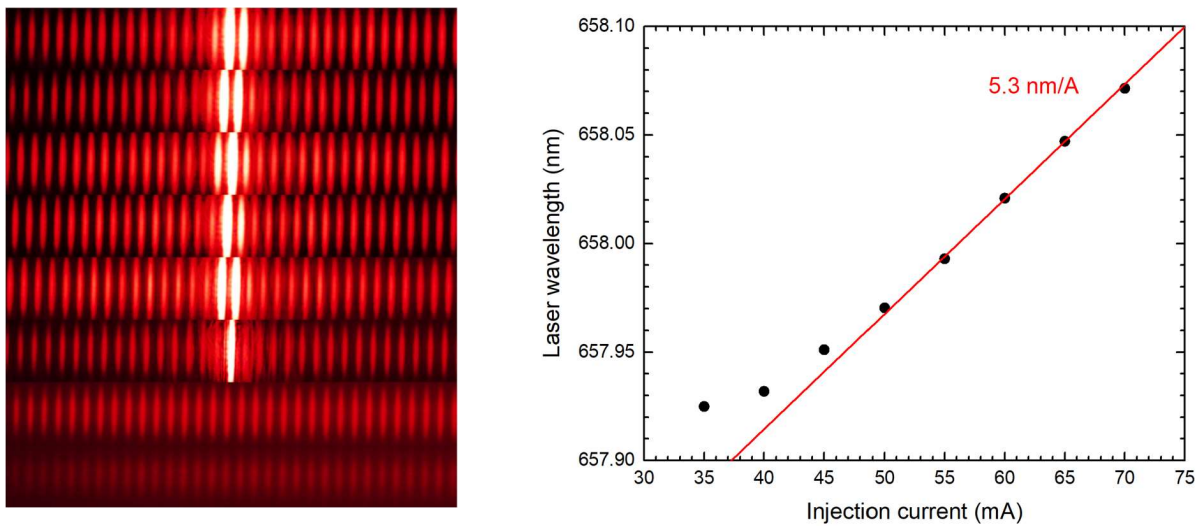

*Figure 15. The composite image on the left show laser spectra using different injection currents, from low (bottom) to high (top), at a fixed laser (heatsink) temperature. The eight current values correspond to the eight data points on the graph on the right. The laser wavelength measurements are for a single cavity mode in the spectra, specifically one that is not one of the dominant lasing modes but a nearby LED-like cavity mode.*

## Optical feedback

We next use a second diffraction grating to retroreflect some of the laser output beam back toward the laser using the optical layout sketched in Figure 16. The retroreflected light acts much like that from an ordinary mirror, forming an "external" laser cavity that augments the "internal" cavity inside the semiconductor medium. This ECDL configuration is illustrated in Figure 5, and the essential physics can be understood from the lower two curves in Figure 6. The *Grating diffraction* curve arises because laser light reflected back into the laser by the grating (which only happens for a specific wavelength) enhances the light buildup in the laser's internal cavity, and this enhances the overall gain at that wavelength. Thus lasing is more likely to happen at this specific wavelength, which is set by the angle of the grating relative to the output laser beam. One can shift the *Grating diffraction* curve simply by rotating the grating, and this "knob" gives us a way to tune the laser wavelength, which is easily seen in the lab.

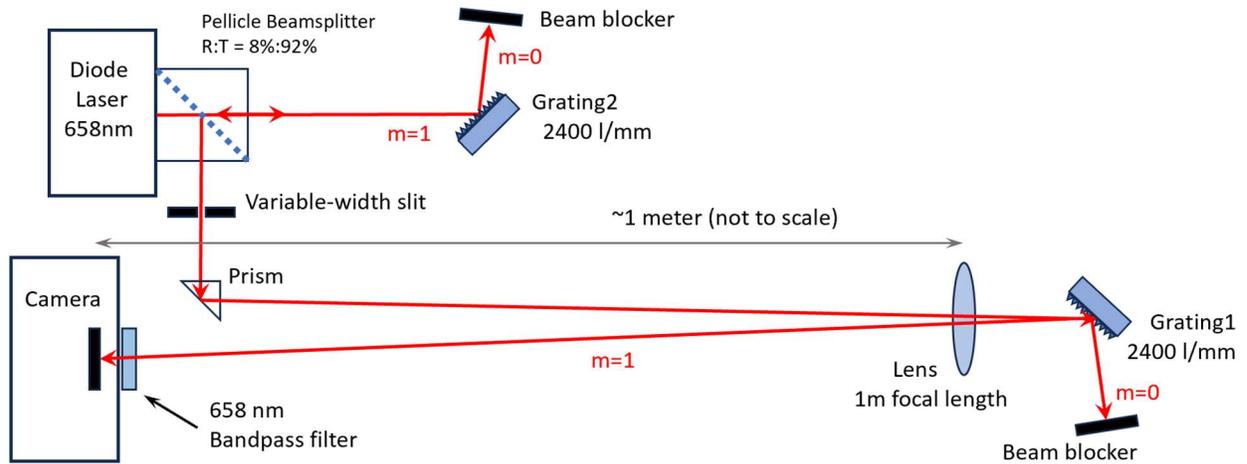

*Figure 16. The layout above builds upon that in Figure 8, adding an additional diffraction grating that uses optical feedback to stabilize the laser output. This External-Cavity Diode Laser (ECDL) configuration allows single-mode laser operation with controllable frequency tuning and laser linewidths below 1 MHz. ECDLs are used extensively in atomic and molecular physics, and in numerous other applications involving tunable-laser spectroscopy.*

In addition, the *External cavity* curve in Figure 6 arises from the additional optical cavity that includes the diode laser, the collimating lens, and the air gap out to the grating. The tilted grating surface acts like a "distributed" mirror that forms the end of the external cavity. Once again, lasing is most likely to occur at the highest generalized gain, which occurs at one of the standing waves in this external cavity.

While the underlying semiconductor and laser physics underlying the ECDL phenomenon is obviously quite complex, observing it in the lab is remarkably easy using the setup in Figure 16. The process begins by setting up Grating2 about 10cm from the laser, and we like the Thorlabs KM100PM mount because it provides a more intuitive adjustment, as seen in Figure 7 (and we use the same mount for Grating1). The grating angle can be coarsely adjusted by sending the laser output beam through a small hole in a white card and then observing the diffracted grating beam on the other side of the card. It is important to make sure the beam is retroreflecting back into the laser quite accurately at this stage. The $m = 1$ beam is about 15% as bright as the main beam, with the other 85% reflecting directly of the grating into the $m = 0$ mode.

To make the fine adjustment of the grating angle, one can use the fact that the laser threshold is lowered by the grating optical feedback, typically by a few mA. If one turns the laser current down to near threshold and adjusts the grating up/down angle, it is straightforward to observe a brightness increase (as seen on the camera monitor) when the diffracted grating beam goes directly back into the semiconductor cavity. At this point optimizing the laser brightness can be used to optimize the grating angle.

Once the grating angle is properly set, increasing the laser current should produce stable single-mode behavior. Moreover, simply adjusting the grating left-right angle will "select" any of the internal cavity modes at will. Figure 17 demonstrates how this looks on the monitor and watching the mode selection yourself in real time as you adjust the grating angle produces something of an empowering feeling. The chaotic frequency behavior seen in Figure 11 is suddenly replaced by the remarkably orderly behavior seen in Figure 17. It's a good teaching moment.

In many ECDL applications, one wants the laser to operate at a specific pre-determined wavelength, for example that corresponding to a particular atomic or molecular transition. To make this happen, all the curves in Figure 6 must combine to form a peak at this wavelength, as described in the figure caption. We like to demonstrate this process in the lab by placing an easily removed "pointer" sticker on the camera monitor, pointing to some random position on the laser spectrum. Students then adjust to grating angle and the laser current to see that this precise frequency (or anywhere else on the spectrum) can easily be produced with an ECDL.

With grating optical feedback in place, we have measured a laser linewidth <5 MHz with the HL6501MG, which is not atypical for ECDL systems at this wavelength. However, the laser frequency is quite jittery because of mechanical noise in the grating mount. With more stable mounting and an added piezoelectric drive of the grating position, ECDLs are well suited for many applications in atomic and molecular spectroscopy.

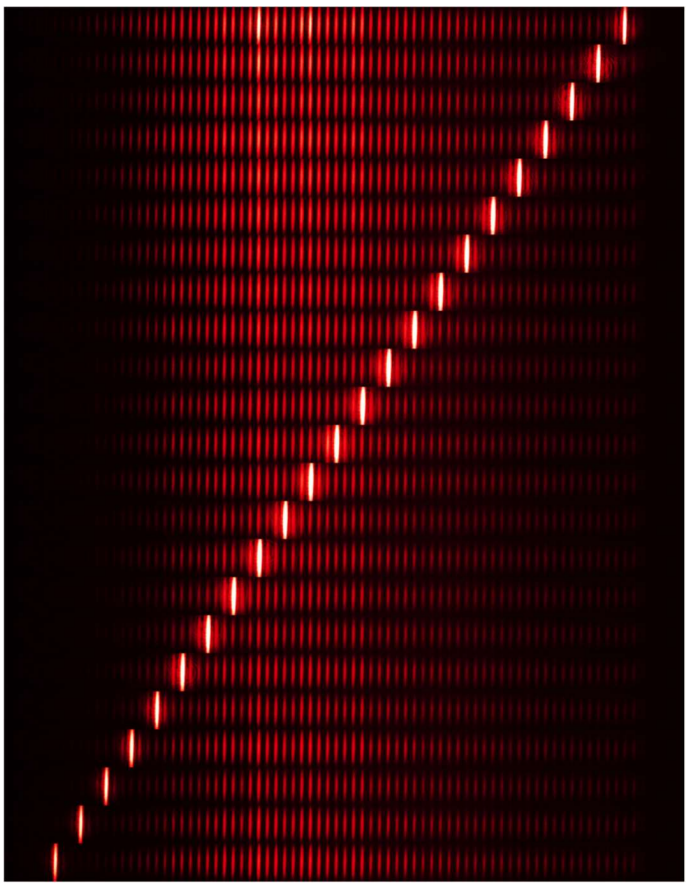

*Figure 17. This composite image illustrates how the laser output changed as the grating angle was changed. The laser temperature and current were fixed, as was the grating up/down angle. Then the grating left/right angle was changed, stopping at every third internal cavity mode to record a spectrum. Any cavity mode can be thus selected, and changing the laser current will move the cavity modes slightly. By changing both the laser left/right angle and the laser current, any predetermined laser frequency can be selected (within the overall range of the diode).*

# Summary

We have found that this experiment works well in our undergraduate teaching lab, providing a hands-on experimental introduction to the physics of semiconductor diode lasers. The hardware is not outrageously expensive, and the modular design should allow periodic upgrades as the better components become commercially available. And with tunable diode lasers playing ever-increasing roles in scientific applications, from laser cooling and trapping of neutral atoms to spectroscopic exploration of other planets, this experiment helps prepare students for working with equipment they might find in graduate research labs and beyond.

# Acknowledgements

This work was supported in part by a generous donation from Beatrice and Sai-Wai Fu to the Physics Teaching Labs at Caltech, together with Caltech's long-standing support of outstanding laboratory instruction across many STEM fields.

*Contact:* For corrections, comments, or just to compare notes, please contact Kenneth G. Libbrecht, *klibbrecht@gmail.com*, or mail to: Mail-stop 264-33 Caltech, Pasadena, CA 91125.

# Appendix 1 – Hardware details

We are still using English-measure optical hardware, so we list parts below from that ecosystem. But metric is clearly preferable if one is just starting out.

## Diode laser

Thorlabs HL6501MG 658nm, 30mW Hitachi laser, $35. The HL6501MG is a 0.65 µm band AlGaInP diode laser with a multi-quantum well structure. Removing an HL6501MG cover and measuring the semiconductor block (Figure 18) gives an overall structure length of 0.787±0.026 mm, which sets an upper bound on the internal cavity length. Combining this with our measurement of $\delta\nu_{cavity} \approx$ 53 GHz for the cavity mode spacing gives an index of refraction of $n \approx 3.6$ near 658 nm, which is typical for AlGaInP diode lasers. Note that absolute maximum rated power of this laser is 35mW. We typically operate the laser below 10mW. At these low powers (and with a good controller that

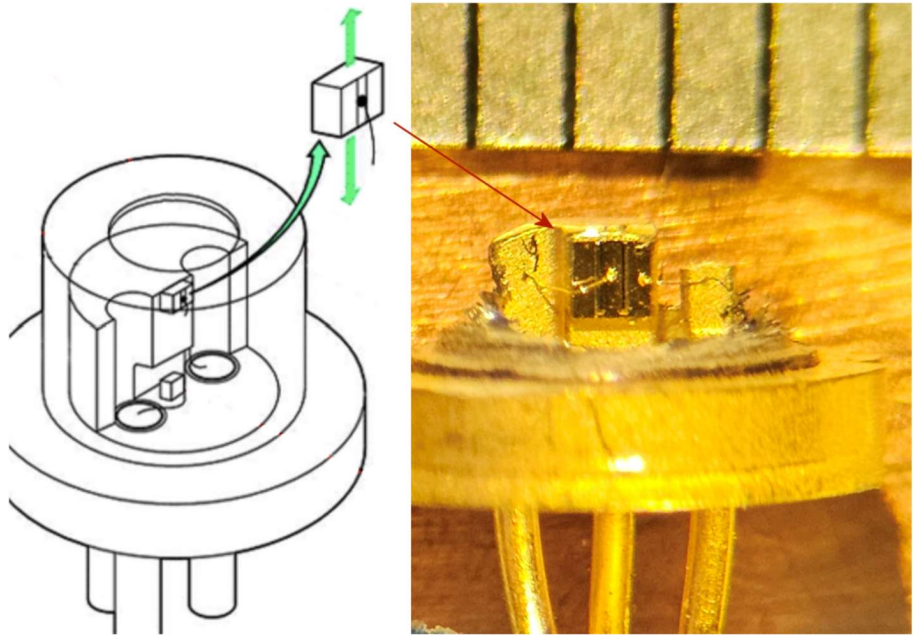

*Figure 18. The sketch on the left shows the basic construction of a diode laser in a can, including a semiconductor block bonded to a large heatsink, with wire-bonded connections between the semiconductor and the external pins. The photo on the right was taken in our lab, showing an HL6501MG laser (after removing the outer can) under a microscope next to a ruler with 1mm divisions. Measuring the image gives a laser cavity length of 0.787±0.026 mm.*

is free of current spikes and static-electric shocks), we find that these diodes last for years in the student lab.

### Laser controller

Thorlabs LTC56A laser diode kit, $3400. This cost-effective solution contains a diode laser mount (supports Ø5.6 mm TO packages with most common pin configurations), a temperature controller (with up to 8W of cooling), a diode current controller (up to 500 mA), and a collimating lens.

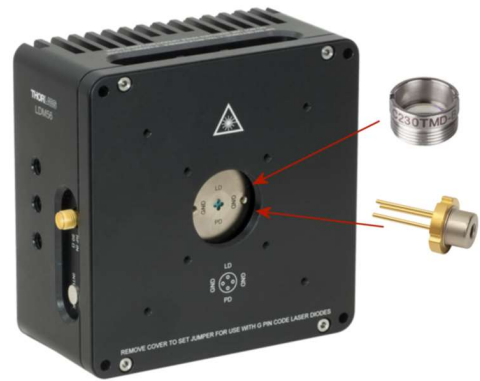

Oddly, the temperature controller in this kit will not display the temperature in degrees but only displays the thermistor resistance value. Fortunately, ChatGPT can easily generate a T/R table based on the thermistor specs (10 k ± 3% at 25 C, NTC, Beta = 3977 K ± 0.75%) in the user manual. Figure 19 shows the temperature response of the diode-laser housing to changes in the temperature set point.

Thorlabs LDMXY flexure mount, $450. This optional item attaches to the LTC56A and allows better collimation of the laser's output beam. The experiment would work without this lens mount,

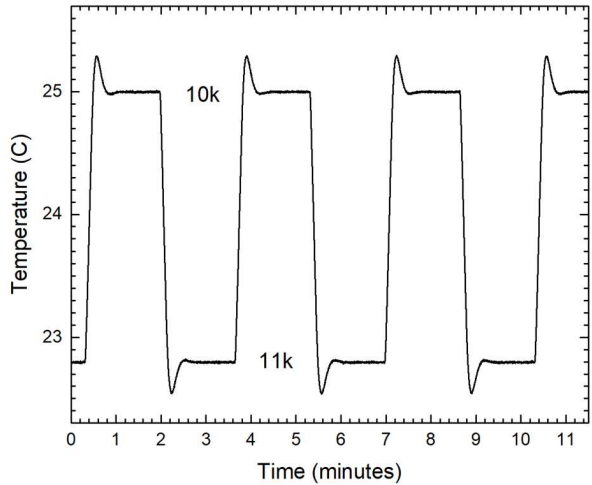

*Figure 19. This plot shows the temperature response of the LTC56A controller, after optimizing the feedback loop. To produce these data, we changed the set point he set point between 10kOhm and 11kOhms using an external square-wave signal generator. The response plotted here shows that the temperature stabilizes nicely in under a minute.*

but it makes the collimation process substantially easier. (Once the collimation is optimized, we place caps over the adjustment knobs to keep students from changing the lens position further.)

## Optical elements

Thorlabs CM1-BP108 mounted pellicle beamsplitter, 8:92 (R:T), $400. Pellicle beamsplitters are preferred in imaging applications (compared to plate or cube beamsplitters), as they minimize light displacement and ghost images. A downside, however, is that the pellicle surface is quite fragile and cannot readily be cleaned. A close look at Figure 7 reveals that we paired our beamsplitter with one SM1CP2M cap and two SM1A64 adaptors to keep dust and prying fingers away from the pellicle beamsplitter. (The fourth port faces the laser mount, with this orientation fixed.)

Thorlabs VA100 adjustable Slit, $330.

Thorlabs MRA10-F01 right-angle prism, $85.

Edmund Optics #43-224 2400 line/mm holographic grating, $150. Two are needed for this experiment. Note that gratings are nearly impossible to clean, so it is best to mount them with some kind of protection to keep fingers (and fingerprints) off the optical surfaces.

Amazon macro close-up lens filter set (any brand), $10. We used a one diopter lens (which has a 1-meter focal length) glued to a post.

Amazon harp-clip style table place-card holders (image on the right), combined with blank white business cards, make excellent inexpensive beam blockers, and they are quite useful for laser-beam alignment.

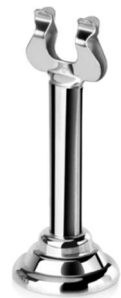

### Camera

Sony A6400 ILCE-6400 mirrorless camera, $800. Just about any camera would work with this experiment. We chose this model because it is relatively inexpensive and includes an HDMI output that can send a live image to a monitor. Note that this "live-view" HDMI feature is not always present in lower priced cameras, but sending the camera output to a 32" monitor is a nice addition to this lab. At its native resolution, the A6400 produces 6000x4000 pixel images with an APS-C size image sensor measuring 23.5 mm × 15.6 mm, thus giving square pixels measuring ~3.91 microns.

Thorlabs FBH660-10 bandpass filter, Ø25 mm, CWL = 660 nm, FWHM = 10 nm, $200. This item is optional, but it does a great job reducing ambient light issues, eliminating any need for light-blocking tubes, which students have little patience for.

### Opto-mechanics

One also needs a variety of posts, post holders, and miscellaneous items to mount the optical elements, and we generally use Thorlabs parts. We especially recommend the Thorlabs UPHA swivel base adapter for student use, as these make alignment easier and more intuitive. All together these parts would cost about $200.

Some kind of optical breadboard is also recommended, such as the Thorlabs MB2448 breadboard that measures 24x48 inches, $1200. A smaller breadboard would work if attached some kind of "outrigger" to hold the far-away lens and grating elements. No vibration isolation is needed for this experiment, so the breadboard can simply be placed on a desk or table for operation.

## Appendix 2 – The internal-cavity mode spacing

I add this Appendix as something of a time capsule demonstrating data analysis in this experiment using AI tools that have only recently burst on the scene (in 2025). While LLMs are still somewhat prone to errors, the latest version of ChatGPT-5 has become good enough to use for everyday physics calculations. It is becoming clear that students will soon be using AI throughout their physics education, as the computer now knows your language so you need not learn to speak its language.

  To explore this brave new pedagogical world, what follows is an analysis of the experiment illustrated Figure 8, specifically using a laser spectrum like that shown in Figure 9b to measure the internal-cavity mode spacing given in Equations (5) and (6). Remarkably, the analysis can be done solely with ChatGPT-5, from start to finish, simply by proving a raw-data image and a series of suitable prompts. How these new AI tools become incorporated into undergraduate physics teaching remains to be seen, but their influence will almost certainly be profound.

# Laser Spectrum Analysis 3

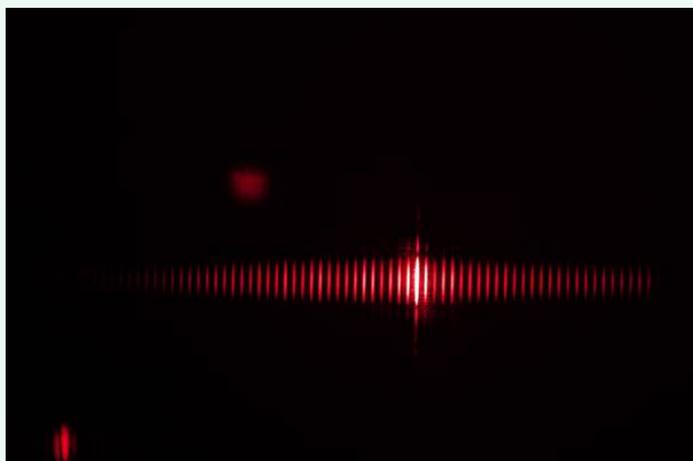

Plot the vertical R intensity profile of this image (where R is the red pixel brightness for RGB pixels), averaging over all horizontal pixels.

**Vertical Red Intensity Profile**

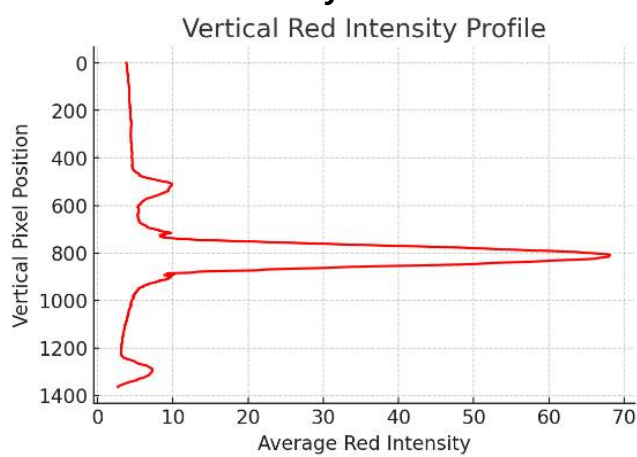

Here's the vertical red intensity profile of your image.

The curve shows the average red pixel brightness across all horizontal pixels, plotted as a function of vertical pixel position. The strong central peak corresponds to the bright diffraction maximum in the image, while the smaller variations trace the weaker fringes and background.

Would you like me to also export this profile as a CSV file for further analysis?

Over what pixel range is this plot above a value of 12? Mark this range on the plot.

**Vertical Red Intensity Profile with Threshold Marked**

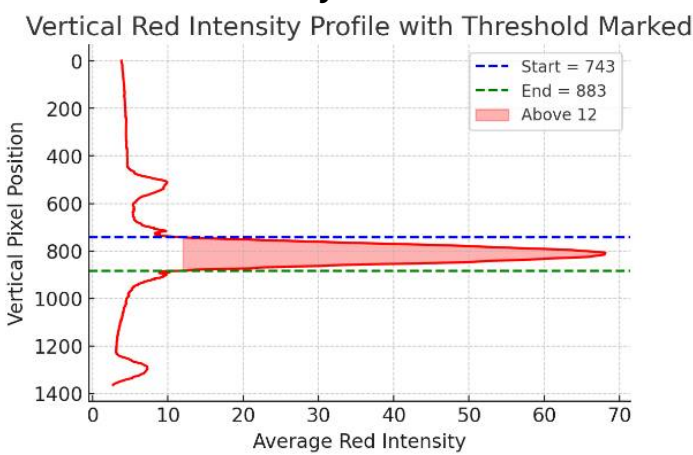

The vertical red intensity stays above **12** between pixel positions **743 and 883**.

On the plot, this range is marked with dashed lines and shaded in red. Would you like me to also compute the full width in pixels (and in physical size if you give me the camera sensor parameters)?

> Average the R pixel values over this vertical range, and plot those pixel values.

**Horizontal Red Intensity Profile (Averaged over rows 743–883)**

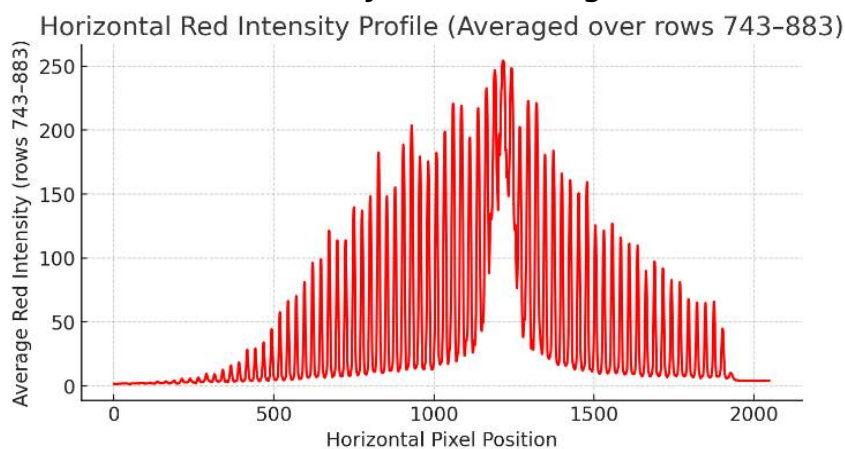

Here's the horizontal red intensity profile, averaged over the vertical range **rows 743–883** where the signal exceeded the threshold.

This plot shows the interference fringes across the horizontal axis, with the central bright maximum and periodic side fringes clearly resolved.

Would you like me to extract the **fringe spacing in pixels** from this curve?

yes

**Detected Fringe Peaks — spacing = 26.00 ± 0.00 px (N=35)**

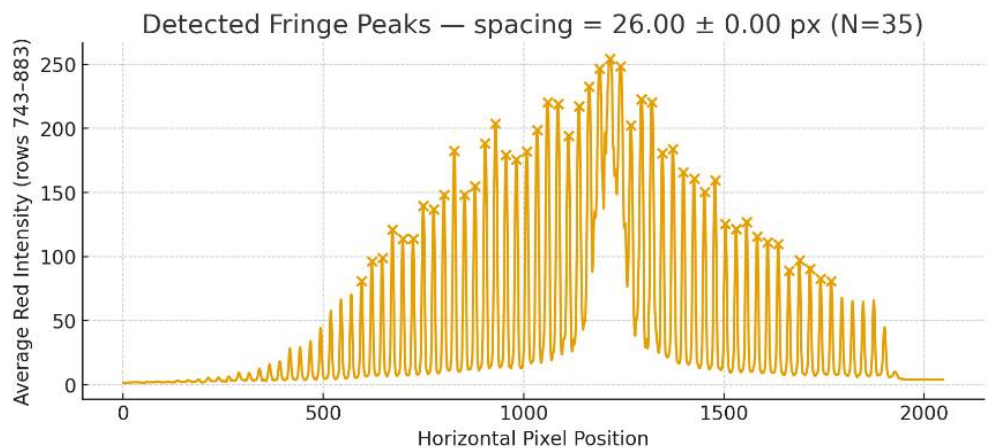

Done. I averaged R over rows **743–883** and found the horizontal fringe spacing from peak-to-peak:

- **Mean spacing: 26.04 px**
- **Std. dev. (1σ): 0.47 px**
- **Number of spacings measured: 45**

I marked all detected peaks on the plot so you can see what was measured.
Want me to export the peak list and spacings to CSV, or convert this spacing to an angular separation using your camera pixel size and screen distance?

The camera sensor has a horizontal size of 23.5 mm. Convert the spacing to physical spacing, and provide a measurement uncertainty in the mean spacing.

Great — using your sensor width **23.5 mm** and the image width **2048 px**, the pixel pitch is:

- **Pixel size:** $23.5\,\mathrm{mm}/2048 = 0.0114746\,\mathrm{mm/px} = 11.4746\,\mu\mathrm{m/px}$

From the averaged profile (rows 743–883), the measured peak-to-peak spacing is:

- **Mean spacing:** $26.044\,\mathrm{px}$
- **Standard deviation of spacings:** $0.475\,\mathrm{px}$ (over $N = 45$ spacings)
- **Uncertainty in the mean (SEM):** $0.071\,\mathrm{px}$

Converted to physical units:

- **Fringe spacing: 0.29885 mm**
- **Uncertainty (SEM): 0.00081 mm** (≈ 0.81 μm)

So the result is:

$$s = (0.29885 \pm 0.00081)\,\text{mm}$$ (mean ± uncertainty in the mean).

> Divide this measurement by 1 meter (the lens focal length) to give the spacing in radians.

The physical fringe spacing of **0.2989 mm** at the sensor, divided by the **1 m focal length**, gives an angular spacing of:

$$\theta = (2.988 \times 10^{-4} \pm 8.12 \times 10^{-7})\,\text{radians}$$

That's about **0.299 mrad ± 0.0008 mrad**.

Would you like me to also convert this to wavelength separation (in nm or GHz) using your earlier grating parameters?

> Grating analysis. Assume a laser wavelength of 658 nm. Assume a diffraction grating with 2400 lines/mm. If the laser beam strikes the grating with an input angle of theta_in, and the output angle theta_out is equal to theta_in, what is theta_in? These angles are relative to the grating normal.

Got it — you're describing the **Littrow configuration** for a diffraction grating.
In Littrow geometry, the incident angle equals the diffracted angle:

$$\theta_{\text{in}} = \theta_{\text{out}} = \theta$$

The grating equation is:

$$m\lambda = d\left(\sin\theta_{\text{in}} + \sin\theta_{\text{out}}\right)$$

For Littrow, this reduces to:

$$m\lambda = 2d\sin\theta$$

## Step 1: Grating spacing

The grating has $2400\,\text{lines/mm}$:

$$d = \frac{1}{2400 \times 10^3}\,\text{m} = 4.167 \times 10^{-7}\,\text{m}$$

## Step 2: Solve for $\theta$

For **first order** diffraction ($m = 1$):

$$\sin\theta = \frac{m\lambda}{2d} = \frac{658 \times 10^{-9}}{2 \times 4.167 \times 10^{-7}}$$

$$\sin\theta \approx 0.789$$

$$\theta \approx \arcsin(0.789) \approx 52.1°$$

✅ **Answer:**

$$\theta_{\text{in}} = \theta_{\text{out}} \approx 52.1°$$

Do you want me to also compute $\frac{d\theta}{d\lambda}$ at this angle for your grating, so you can compare with your measured 0.299 mrad fringe spacing?

> Assume that theta_in is fixed at 52.6 degrees. What is d(theta_out)/d(lambda) in this case?

Perfect — now we hold $\theta_{\text{in}}$ fixed and solve for how the **output angle** changes with wavelength.

## Step 1: Grating equation

General grating equation:

$$m\lambda = d\bigl(\sin\theta_{\text{in}} + \sin\theta_{\text{out}}\bigr)$$

Here:

- $m = 1$ (first order),
- $d = 4.167 \times 10^{-7}$ m,
- $\lambda = 658\,\text{nm} = 6.58 \times 10^{-7}$ m,
- $\theta_{\text{in}} = 52.6°$.

So:

$$\sin\theta_{\text{out}} = \frac{m\lambda}{d} - \sin\theta_{\text{in}}$$

## Step 2: Differentiate

Differentiate w.r.t. $\lambda$:

$$\cos\theta_{\text{out}} \cdot \frac{d\theta_{\text{out}}}{d\lambda} = \frac{m}{d}$$

So:

$$\frac{d\theta_{\text{out}}}{d\lambda} = \frac{m}{d \cos \theta_{\text{out}}}$$

## Step 3: Compute $\theta_{\text{out}}$

$$\sin \theta_{\text{out}} = \frac{6.58 \times 10^{-7}}{4.167 \times 10^{-7}} - \sin(52.6°)$$

$$= 1.578 - 0.793 \approx 0.785$$

$$\theta_{\text{out}} = \arcsin(0.785) \approx 51.8°$$

## Step 4: Evaluate derivative

$$\cos \theta_{\text{out}} = \cos(51.8°) \approx 0.617$$

$$\frac{d\theta_{\text{out}}}{d\lambda} = \frac{1}{(4.167 \times 10^{-7})(0.617)} \approx 3.88 \times 10^6 \, \text{rad/m}$$

Convert to more practical units:

- Per **nm** (1 nm = $10^{-9}$ m):

$$\frac{d\theta_{\text{out}}}{d\lambda} \approx 3.88 \times 10^{-3} \, \text{rad/nm}$$

$$= 3.88 \, \text{mrad/nm}$$

✅ **Answer:**

$$\frac{d\theta_{\text{out}}}{d\lambda} \approx 3.88 \, \text{mrad/nm at } \theta_{\text{in}} = 52.6°$$

Would you like me to now combine this with your **measured angular spacing** (0.299 mrad) to compute the corresponding wavelength separation (in nm and GHz)?

> Assuming the angular spacing you calculated above, what is the wavelength separation between the peaks in nm? Express this in GHz as well.

Using your measured angular fringe spacing (0.299 mrad) and the grating derivative:

- **Wavelength separation:**

$$\Delta\lambda = (0.0770 \pm 0.0002) \, \text{nm}$$

- **Frequency separation:**

$$\Delta\nu = (53.37 \pm 0.15)\,\text{GHz}$$

Would you like me to also compare these theoretical values with what you'd expect from your camera-based measurement (i.e., pixel spacing and geometry), as a consistency check?

> These peaks are from the internal cavity of a diode laser. If the cavity length is 0.787 mm, what is the index of refraction of the semiconductor material?

Great — now we can connect your measured **mode spacing** to the **cavity parameters**.

## Step 1: Mode spacing formula

For a Fabry–Pérot cavity:

$$\Delta\nu = \frac{c}{2nL}$$

where:

- $c$ = speed of light in vacuum,
- $n$ = refractive index of the medium inside the cavity,
- $L$ = cavity length,
- $\Delta\nu$ = longitudinal mode spacing (frequency separation).

## Step 2: Solve for $n$

$$n = \frac{c}{2L\,\Delta\nu}$$

We already found:

- $\Delta\nu = 53.37 \times 10^9\,\text{Hz}$,
- $L = 0.787\,\text{mm} = 7.87 \times 10^{-4}\,\text{m}$,
- $c = 3.00 \times 10^8\,\text{m/s}$.

$$n = \frac{3.00 \times 10^8}{2 \times (7.87 \times 10^{-4}) \times (5.337 \times 10^{10})}$$

## Step 3: Calculate

$$n \approx 3.60$$

✅ **Answer:**
The effective index of refraction of the diode's semiconductor cavity is about:

$$n \approx 3.6$$

That's in line with typical values for GaAs- and AlGaInP-based diode lasers (≈ 3.3–4.0).

Would you like me to also propagate your uncertainty from the measured fringe spacing (± 0.15 GHz) to give an uncertainty in $n$?